\documentclass[prb,twocolumn,preprintnumbers,amsmath,amssymb,superscriptaddress,nofootinbib]{revtex4}
\usepackage{amsmath,amssymb}
\usepackage{epsfig,graphicx}
\usepackage{subfigure}
\usepackage{graphicx}
\usepackage{rotating}
\usepackage{bm}
\usepackage{color}
\newcommand{\be}{\begin{equation}}
\newcommand{\ee}{\end{equation}}
\def\bea{\begin{eqnarray}}
\def\eea{\end{eqnarray}}

\long\def\symbolfootnote[#1]#2{\begingroup%
\def\thefootnote{\fnsymbol{footnote}}\footnote[#1]{#2}\endgroup}

\begin{document}

\author{Steven~A. Abel}
\affiliation{Institute for Particle Physics Phenomenology, University of Durham, Durham, DH1 3LE, UK}
\author{John Ellis}
\affiliation{Theory Division, Physics Department, CERN, CH-1211 Geneva 23, Switzerland}
\author{Joerg Jaeckel}
\affiliation{Institute for Particle Physics Phenomenology, University of Durham, Durham, DH1 3LE, UK}
\author{Valentin~V. Khoze}
\affiliation{Institute for Particle Physics Phenomenology, University of Durham, Durham, DH1 3LE, UK}

\title{
\vspace*{-0.7cm}\hfill{\small CERN-PH-TH/2008-154}\\[-0.2cm]\hfill{\small IPPP/08/50}\\[-0.2cm]\hfill{\small DCPT/08/100}\\
\vspace{20pt}
\Large{\textbf{Will the LHC Look into the Fate of the
Universe?\footnote{This is a brief review targeted primarily at a non-expert audience.}}}}

%onecolumngrid
\begin{abstract}
%{\textbf{
The LHC will
probe the nature of the vacuum that determines the properties of
particles and the forces between them. Of particular importance is
the fact that our current theories allow the Universe to be trapped
in a metastable vacuum, which may decay in the distant future,
changing the nature of matter. This could be the case in the
Standard Model if the LHC finds the Higgs boson to be light.
Supersymmetry is one favoured extension of the Standard Model which
one might invoke to try to avoid such instability. However, many
supersymmetric models are also condemned to vacuum decay for
different reasons. The LHC will be able to distinguish between
different supersymmetric models, thereby testing the stability of
the vacuum, and foretelling  the fate of the Universe.
%}}
\end{abstract}
%\vspace*{0.3cm}
%\twocolumngrid
%\bigskip

\maketitle

According to quantum physics, what we see as particles are
really excitations out of
the ``vacuum''. The precise configuration
of the vacuum, in particular its symmetries,
are determined by the fact that the Universe
seeks to minimize its energy. The symmetries of the
vacuum then determine the fundamental forces acting
on matter, i.e., the three ``gauge'' forces (electromagnetic, weak and strong) and
gravity.

\begin{figure}
\includegraphics[width=.45\textwidth]{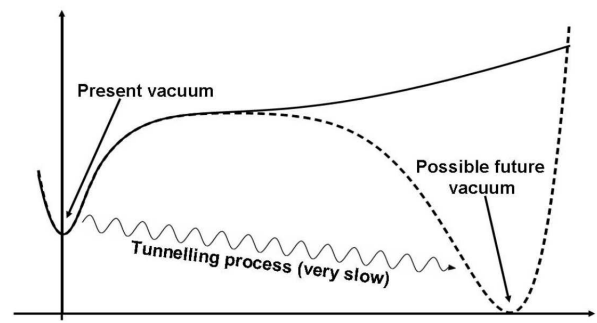}
\caption{\label{fig:potential}\it
Sketch of the potential energy.
The solid line shows the potential in a theory with only one stable vacuum
(i.e., a single minimum of the energy). The dashed
line represents the potential in a slightly modified theory
where there is an additional minimum of lower energy, so that the
first ``false" minimum is only metastable. In the latter theory,  the
Universe can undergo a phase transition where it eventually tunnels
through the barrier and ends up in the ``true" minimum of lower energy (wavy
arrow). Since particles correspond to small excitations around a minimum, changing the minimum itself would correspond to a drastic change in particle properties.
However, depending on the height and width of the barrier,
this process can be extremely slow and may need many, many billions
of years. Although the LHC will probe only a
small region around
the present minimum (and therefore can never trigger a crossing of the barrier),
 in many cases we may
nevertheless be able to tell whether we live in a false vacuum.}
\end{figure}

Although the laws of physics appear to be constant today,
the vacuum need not be inert on cosmological timescales:
it can occasionally undergo dramatic shifts as, for example,
quantum tunnelling of the Universe into a vacuum configuration with
lower energy (cf. Fig.~\ref{fig:potential}).
The unstable but very long-lived vacua are called
``metastable''.
The eventual tunneling transition
results in a new vacuum which
has, in general, different symmetries and hence different fundamental
forces. Vacuum shifts of this type are
``phase transitions'' -- analogous to the
boiling or freezing of water.
In the particle physics context they are
also known as ``vacuum decay''.

One such symmetry-changing phase transition is thought
to have happened shortly after the Big Bang.
It is almost certain that at this stage there was a vacuum shift that
``broke'' the symmetry underlying the weak nuclear force causing it
to freeze out so that
it plays very little role in our everyday life.
Earlier on, another phase transition may have
been responsible for the present relative strength of the strong nuclear
force,
and an even earlier phase transition
may have caused the emergence of space and time themselves.
Has the vacuum now settled down, or will it change again?
As we shall see, in
our current best guesses about
the next level of particle physics, the eventual
decay of the vacuum is a very reasonable
possibility.
The LHC collider now being commissioned at CERN
is designed to reach unprecedented high energy scales. It
will be the first accelerator to probe directly the nature
of the vacuum, and hence the possible fate of the Universe.

Within our current paradigm of particle physics,
the Standard Model, vacuum decay could occur if the Higgs boson,
the quantum excitation of the vacuum that manifests the breaking
of its symmetry, is relatively light~\cite{Frampton,radcorr}.
A relatively light Higgs is even favoured indirectly by data from lower-energy accelerators~\cite{EWWG}.
Discovering the Higgs boson and accurately measuring its mass is
one of the primary objectives of the LHC~\cite{LHC}.
Hence, if the Standard Model is all there is, the LHC will tell us about
the stability of its vacuum.

However, many expect more than just the Standard Model to be found at the LHC.
One promising extension of the Standard Model is superymmetry,
and indeed this seemed a promising way to stabilize the vacuum~\cite{ER}.
However, recent developments~\cite{Intriligator:2006dd,Intriligator:2007py}
suggest instead that supersymmetry may actually condemn the Universe to vacuum
decay (for a precursor, see~\cite{Ellis:1982vi}).
Again, the LHC may tell us whether Nature has this fate in store.

Supersymmetry~\cite{susy} is a symmetry that relates elementary particles of integer
and half-integer spin, known as bosons and fermions, respectively.
The Standard Model is not supersymmetric. Although it contains examples of
both, bosons (e.g., gauge bosons such as photons which carry force,
and the Higgs boson that gives particles their masses) and fermions (e.g., matter particles
such as quarks and electrons) they are not related by any symmetry.
In a supersymmetric extension of the Standard
Model, every particle would acquire a superpartner whose properties
such as charge and mass are exactly the same, and which differ
only in their spin.
For example, gauge bosons are accompanied by fermions called
gauginos, and quarks are accompanied by bosons called squarks.

Supersymmetry provides a solution to profound conceptional problems
of particle physics by taming certain infinities in the theory, and offers
practical advantages such as a candidate for the dark matter that
appears to be cluttering up the Universe~\cite{Zwicky:1933gu}.
For these reasons, many physicists expect it to play a significant role in
Nature. However, at currently accessible energies no superpartners have
been found. Our best guess is that supersymmetry is
itself a broken symmetry much like the broken symmetry of the
weak nuclear force -- this could result in higher superpartner masses
and explain why they have so far evaded detection, but would leave intact
the nice mathematical properties which made supersymmetry so attractive
in the first place. The LHC may well discover supersymmetry~\cite{LHC}, in
particular if it provides the dark matter, as well as
a light Higgs boson.

Even within the simplest supersymmetric version of the Standard Model
(commonly called the Minimal Supersymmetric
extension of the Standard Model or MSSM),
a particular choice of supersymmetry
breaking pattern can lead to an unstable vacuum. Generally, around
half of the available parameter
space of superpartner masses leads to metastable vacua. In
this case the endpoint of the eventual decay would be a new
vacuum with very little symmetry and hence practically
no forces acting at all except gravity~\cite{Casas:1995pd}.
The change in physics caused by these
transitions would be drastic -- in the new vacuum, atoms and nuclei
would fall apart, and the Universe
would become a soup of heavy, decoupled particles.

So far, metastability of the vacuum seems to be just a quirk of the
theory, which could, in principle, be avoided in a sizeable part of
parameter space. Recent developments however suggest that in large
classes of supersymmetric theories, metastability is essentially
inevitable, and in fact the consistency of the theory requires it.
The arguments hinge on one niggling issue with supersymmetry: it is
hard (in a mathematical sense) to break it. As we shall see, if the
breaking of supersymmetry is realised in one favoured way it is
practically guaranteed that, in addition to the vacuum in which the
Universe currently resides, there is another state of lower energy -
often called the ``true'' vacuum - in which supersymmetry is
unbroken (the present metastable vacuum is often referred to as a
``false'' vacuum). Everyday physics in false and true vacua is hard
to distinguish, and in particular all the matter in the Universe we
see today could very well be composed of elementary particles which
are quantum excitations over a false vacuum, that is susceptible to
decay to the true vacuum in which supersymmetry is an exact symmetry
of Nature. Moreover, the alternative scenarios for supersymmetry
breaking {\em also} offer future vacuum decay as a possibility, if
not an inevitability. The LHC cannot trigger this change in the
vacuum\symbolfootnote[2]{If they could, cosmic-ray collisions would
already have done so long ago~\cite{Voloshin}.}, but it can serve as
a crystal ball that reveals the fate of the Universe.

The recent developments involve a rather subtle web of theoretical
and experimental arguments, and so before we
describe them we should (perhaps to cheer the reader up a little)
discuss the timescales on which our present vacuum would decay, as
indicated by the wavy line in Fig.~1. There is
very firm evidence that the laws of physics have
been constant since
the first few minutes after the Big Bang. Thus, unless we are
rather unlucky, one
would expect that the timescale would be
at least billions of years. However, we can do better than
that: quantum mechanics allows us to compute the lifetimes of
false vacua.
The physics underlying false vacuum decay was elucidated in a beautiful series of
papers by Coleman and de Luccia~\cite{Coleman:1980aw}.
They found that a false vacuum decays by creating huge instantaneous
``lumps'' of particles, called (appropriately enough) instantons, on
which bubbles of the new vacuum nucleate. The effect of such a lump,
and hence the rate of decay, increases with the strength of the
interactions in the theory. For example, the simplest instantons,
which are comprised of only gauge bosons, allow tunneling between
different ``gauge'' vacua with a rate proportional to an
exponentially small factor, $e^{-\frac{8\pi^2}{g^2}}\ll 1$, where
$g$ is the interaction strength. The important point is that it is
the self-interaction of the gauge bosons (i.e. the fact that $g$ is
not zero) which allows transitions between  different vacua. The
decay of a false vacuum is catalyzed by analogous but more
complicated instantons which interpolate across the energy barrier
separating true and false vacua, and the forms of the energy
barriers (in particular their heights and widths, cf.
Fig.~\ref{fig:potential})
determine the typical decay time.
The probability of vacuum decay is
always exponentially suppressed in this manner, and
in all realistic models, this makes
the supersymmetry-breaking false vacuum very long-lived, with a life
expectancy much longer than the age of the Universe.

Let us now return to why we expect that
metastability is unavoidable in certain well-defined
and testable scenarios.
This requires more information on how supersymmetry may be broken.
In order to fit in with our current (lack of) observations,
supersymmetry is almost certainly broken by what is known as
a ``hidden sector''. This is a part of the theory that interacts extremely
weakly with the particles of the supersymmetric Standard Model -- called in this
context the ``visible sector''. The supersymmetry breaking that ends up
in the visible sector is filtered and weakened through these interactions, a
process known as mediation. The models can be classified according
to how the supersymmetry is broken in the hidden sector and how this
breaking is mediated to the visible sector.

For example, the supersymmetry-breaking hidden sector could interact
with the visible sector through gravitational interactions alone, the option
known as gravity mediation.
Since every particle interacts with gravity, this is the weakest sort of
mediation one could imagine, and in these models the scale of
supersymmetry breaking in the hidden sector has to be
very high -- the energy scales involved are roughly $10^{11}$~GeV (remember
that the proton mass is about 1~GeV).
An alternative scenario
is that the supersymmetry-breaking hidden sector interacts with the visible sector
through the other fundamental
forces as well, i.e., the electroweak and strong gauge interactions.
In this case the mediation is stronger and the scale of supersymmetry breaking
in the hidden sector is correspondingly lower, typically $10^{5-7}$~GeV.

What about the way that supersymmetry is broken? Either matter fields dominate the
supersymmetry breaking~\cite{O'R}, the option known as $F$-term breaking, or gauge
fields dominate~\cite{FI}, the option known as $D$-term breaking.
An important difference is that $F$-term breaking is
calculable in the sense that we have full mathematical control,
whereas $D$-term breaking is rather more difficult to handle -- there are
unknown factors that affect the size of supersymmetry breaking or indeed
whether supersymmetry is broken at all. In order to be able to make firm conclusions here,
therefore, we consider $F$-term breaking.

\begin{figure}
\includegraphics[width=.5\textwidth,angle=0]{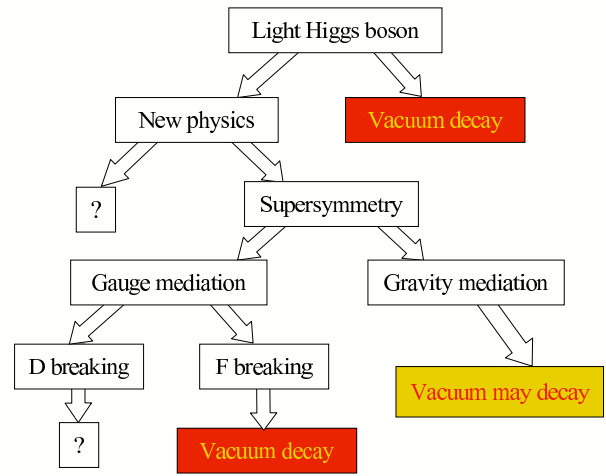}
\caption{\label{fig:metastability}\it The LHC may
reveal the fate of the Universe by discovering a light Higgs boson
and/or supersymmetry. In calculable gauge-mediated models of
supersymmetry breaking the present vacuum is necessarily unstable,
and the same may be true in gravity-mediated models. }
\end{figure}

We represent the situation schematically in Fig.~\ref{fig:metastability},
where
the main classes of supersymmetric models are depicted.
The class of models that we consider primarily are the
gauge-mediated models with $F$-term breaking --
the scenario we refer to as `calculable gauge mediation' (CGM).
It is in these models that metastability is
generically unavoidable as we shall now see.
The LHC has the capability to tell us whether Nature realizes this option.
Metastability may occur in the other scenarios as well, but
may be avoidable, by for example a judicious choice of parameters. As we discuss
at the end, in this case the LHC can also reveal whether Nature has chosen
metastability.

The seeds of the inevitability of metastability in this class of models lie in an
important theorem due to Nelson and Seiberg~\cite{Nelson:1993nf}, who
identified a \emph{necessary and sufficient} condition for $F$-term supersymmetry breaking.
Called $R$-symmetry, this condition is a generalisation of the familiar rotations in space
that is unique to supersymmetric theories. The problem arises because in
$R$-symmetric theories the supersymmetric partners of the gauge bosons -
called gauginos - must be massless, in conflict with experiments, which require
$m_{gaugino} \gtrsim 100$~GeV.
The dilemma is that non-vanishing gaugino masses require both supersymmetry breaking
{\em and} $R$-symmetry
breaking, but Nelson and Seiberg tell us that these two requirements are
mutually exclusive. How to get around it?

There are two logical possibilities.
One is to include in the theory a small, controlled amount of $R$-symmetry breaking.
More precisely, the Lagrangian function, which defines all masses and
interactions of the theory, would be of the form
\begin{equation}
\label{lagrangian}
\mathcal{L}=\mathcal{L}_{R}+\varepsilon\mathcal{L}_{R-breaking},
\end{equation}
where $\mathcal{L}_{R}$ describes a theory which preserves $R$-symmetry and breaks supersymmetry,
whereas $\mathcal{L}_{R-breaking}$ breaks $R$-symmetry, and
$\varepsilon$ is our small control parameter.
When $\varepsilon=0$, the lowest-energy (ground) state breaks supersymmetry,
and there is no supersymmetric vacuum at all (solid line in Fig.~\ref{fig:potential}),
but the gauginos are massless.
However, with a small
$\varepsilon \ne 0$, $R$-symmetry is broken explicitly.
In this case, the Nelson-Seiberg theorem requires that a
supersymmetry-preserving vacuum
appears in addition to the
supersymmetry-breaking one (dashed line in Fig.~\ref{fig:potential}),
since the full theory breaks $R$-symmetry.
It is a general consequence of supersymmetry that any supersymmetric vacuum must be the state
of lowest energy. Hence, the non-supersymmetric vacuum must be
 metastable. However, it is important to note that the two vacua are separated by
a distance that goes to infinity as $\varepsilon\rightarrow 0$ (in Fig.~\ref{fig:potential} the new minimum moves further and further to the right).
As the control parameter
$\varepsilon \to 0$, the decay rate of our false vacuum becomes
exponentially longer and longer.

The second possible way to obtain non-vanishing
gaugino masses is for the vacuum itself to break
the $R$-symmetry -- a possibility known as spontaneous breaking.
With spontaneous breaking, the whole theory still obeys
the symmetry, but the effective physics we see in the
symmetry breaking vacuum does not.
Spontaneous (rather than explicit) breaking of $R$-symmetry does not introduce
new
supersymmetry preserving minima, and does not by itself make the supersymmetry
breaking vacuum metastable.
In particular we do not need to introduce and explain the
origins of a very small parameter $\varepsilon$, as we had
to with explicit breaking.
At the same time, gauginos acquire masses proportional to the scale of
spontaneous $R$-breaking.

Superficially then, it looks as if one might be able to avoid metastability. Alas,
spontaneous symmetry breaking involves some subtleties: we begin with a
symmetric theory and choose a vacuum that breaks it. But, since the
original theory had a symmetry, so must the set of choices of possible vacua. In
other words, there is a degeneracy of vacua all with the same energy, corresponding precisely to the
symmetry we are breaking. Since small
fluctuations in the choice of vacuum do not cost any energy, there must be
a new massless particle -- the Goldstone mode, that reflects this symmetry.
In the case of spontaneously-broken $R$-symmetry, this particle is called the $R$-axion.
In order to avoid astrophysical and experimental bounds, the $R$-axion
must also acquire a mass, although the lower bounds on its mass are much weaker than
those on the gaugino mass: $m_{R-{\rm axion}} \gtrsim 100$~MeV, and
therefore easier to fulfill. Nevertheless, its mass means that the original
$R$-symmetry must itself be
explicitly broken by very small effects,
and according to the earlier arguments,
this again implies that the vacuum is metastable. In this case, however,
the gaugino mass is divorced from the size of the \emph{explicit} $R$-breaking parameter
$\varepsilon$, which now determines the $R$-axion mass instead.
This exhausts the logical possibilities and shows that massive gauginos
and massive $R$-axions imply metastability.

An invaluable contribution to these arguments was made by the
recent papers of Intriligator, Seiberg and Shih (ISS) \cite{Intriligator:2006dd,Intriligator:2007py}.
In particular, the question that had previously been unanswered
was how to generate a Lagrangian
of the form (1). ISS discovered an extremely simple and beautiful class of supersymmetric
models that generate dynamically a small $R$-breaking term of the required type by
quantum effects, and hence lead to a long-lived metastable vacuum.
The Nelson-Seiberg theorem manifests itself in a
truly wonderful way in these theories: the classical theory (i.e.,
before we add in quantum effects) has an exact $R$-symmetry.
However, the quantum theory does not preserve the $R$-symmetry - the $R$-symmetry
is said to be ``anomalous'', which guarantees that small effects of the type
$\varepsilon\mathcal{L}_{R-breaking}$ will appear. In the ISS models,
$\varepsilon$ is a naturally small parameter, because
it too is generated by
instanton-like effects and hence is proportional to the factor $e^{-8\pi^2/g^2} \ll 1$.

This breakthrough has led to a burst of activity
building gauge-mediated models incorporating the ISS models as hidden sectors.
The complementary explicit and spontaneous approaches to model-building
were successfully incorporated with a few twists.
In the first approach, the explicit $R$-breaking of the ISS
models was not able to generate gaugino masses, so a second source
of $R$-breaking was required.
However, the smallness of this second term -
necessary for the longevity of the metastable vacuum,
turned out to be guaranteed within the ISS models if the $R$-symmetry-breaking
effects were generated at a very large energy scale,
e.g., the Planck scale~\cite{Murayama:2006yf}.

In the second approach, the gauginos are already massive and, as we discussed above,
the job of the explicit $R$-breaking is merely to give the $R$-axion a small mass
$m_{axion}\gtrsim$100 MeV. The controlled quantum effects within
all models of the ISS type are sufficient to do this, and
remarkably simple
versions of the ISS model could be found that led to the required
spontaneous $R$ breaking~\cite{Abel:2007jx,Abel:2007nr},
so that gauginos
receive sufficiently large masses $m_{\rm gaugino} \gtrsim 100$~GeV.
These are explicit, credible CGM models with metastable vacua.
The LHC will be able to produce gauginos weighing an order of
magnitude more than the present lower limit~\cite{LHC}, offering a good prospect
of testing such metastable CGM scenarios.

We have argued that, in such CGM scenarios of supersymmetry
breaking, metastability is generically unavoidable because gaugino
and $R$-axion masses must both be non-zero. Until now, we have not
addressed the problem of the cosmological constant: global
supersymmetry breaking {\` a} la CGM always generates a large vacuum energy.
This makes a contribution to the cosmological constant that is much larger than the observed
tiny value.

This contribution
can in principle be compensated
in a ``supergravity'' theory, i.e., a theory combining supersymmetry with gravity,
which can easily generate an additional negative contribution to the vacuum energy.  Adding this
contribution would not change our conclusions about the metastability of the vacuum in
CGM models. However, supergravity does offer the alternative possibility of using
gravitational-strength interactions to mediate supersymmetry breaking, as illustrated in Fig.~2.
There is no theorem in such gravity-mediated models that our present vacuum is necessarily
unstable. However, this is still a generic possibility~\cite{Casas:1995pd},
and the cosmology of such metastable
gravity-mediated scenarios was
discussed recently\cite{EGLOS}. These scenarios
can also be probed by the LHC, through measurements of the spectrum of
supersymmetric particles, should they be discovered.

Before closing, we address one question that may have been nagging the reader: if our
present vacuum is a ``false" one, how did the Universe arrive in
such a metastable minimum? Why did it not
start directly in the ``true" stable vacuum?
In the models discussed above the reason is that the early Universe was
(presumably) very hot. At high temperatures, what later became
the metastable vacuum is preferred by entropy, and the Universe was automatically driven into it.
Later, as the Universe cooled, it got trapped in the metastable
state~\cite{Abel:2006cr,Craig:2006kx,Fischler:2006xh} (for a precursor, see~\cite{Ellis:1982vi}).

{\emph{In conclusion,}} we have argued that the present ground state of
the Universe may well be temporary, and that it may ultimately decay into an
energetically more favourable one. This could arise in the Standard Model, for example, if
the Higgs boson is light. One way to avoid this would be to postulate an extension
of the Standard Model to include supersymmetry. However,
the metastability of our present vacuum is also unavoidable in a generic class
of supersymmetric theories. In
either case, the LHC may be able to indicate
whether Nature is metastable
by for example discovering the Higgs boson and
measuring its properties, or by discovering superpartners
and measuring their masses.
By studying the nature of the
vacuum, the LHC will provide a unique window on the fate of the
universe.

\end{document}